\documentclass[letterpaper, 10 pt, conference]{ieeeconf}  
\IEEEoverridecommandlockouts                              

\usepackage{xspace,empheq,fancybox,amsmath,amssymb,graphicx,epstopdf,epsfig,subfigure,syntonly,times,amsthm} 
\usepackage{psfrag,color,bm,array,cite}
\usepackage{url,cite,footnote,xspace,syntonly,algorithm,algorithmic}
\usepackage{verbatim,multirow}
\usepackage[T1]{fontenc}
\usepackage{mathtools}
\usepackage{amsmath,amsfonts,amsthm,bm} 
\usepackage{graphicx}
\usepackage{mwe}
\usepackage{caption}
\usepackage{cite}
\usepackage{amsmath,amssymb,amsfonts}
\usepackage{algorithmic}
\usepackage{graphicx}
\usepackage{textcomp}
\usepackage{xcolor}
\usepackage{multicol, blindtext}
\usepackage{xspace,empheq,fancybox,amsmath,amssymb,graphicx,epstopdf,epsfig,syntonly,times,amsthm} \usepackage{psfrag,color,bm,array,cite}
\usepackage{url,cite,footnote,xspace,syntonly,algorithm,algorithmic}
\usepackage{verbatim,multirow}
\usepackage[T1]{fontenc}
\usepackage{mathtools}
\usepackage{subfig}
\usepackage{amsmath,amsfonts,amsthm,bm}
\usepackage{graphicx}
\usepackage{mwe}
\usepackage{caption}
\usepackage{stfloats}
\usepackage{amsfonts}
\usepackage{stackengine}
\usepackage{lipsum}
\usepackage{bbm, dsfont}

\usepackage[T1]{fontenc}
\usepackage{dsfont}

\newtheorem{corollary}{Corollary}

\newtheorem{proposition}{\hspace{0pt}\bf Proposition}

\newtheorem{remark}{\bfseries Remark}

\usepackage{amsthm}

\newcolumntype{P}[1]{>{\centering\arraybackslash}p{#1}}

\newcommand*{\myov}[1]{{\overbracket[0.65pt][-1pt]{#1}}}
\renewcommand{\bar}[1]{\myov{#1}}

\title{\LARGE \bf
Machine Learning for Fairness-Aware Load Shedding: \\ A Real-Time Solution via Identifying Binding Constraints}

\author{Yuqi Zhou, Joseph Severino, Sanjana Vijayshankar, Juliette Ugirumurera, and Jibo Sanyal
\thanks{\protect\rule{0pt}{3mm} 
This work was authored by the National Renewable Energy Laboratory, operated by Alliance for Sustainable Energy, LLC, for the U.S. Department of Energy (DOE). The views expressed in the article do not necessarily represent the views of the DOE or the U.S. Government. The U.S. Government retains and the publisher, by accepting the article for publication, acknowledges that the U.S. Government retains a nonexclusive, paid-up, irrevocable, worldwide license to publish or reproduce the published form of this work, or allow others to do so, for U.S. Government purposes.
}
\thanks{The authors are with the National Renewable Energy Laboratory, Golden, CO, 80401, USA. Emails:
        {\tt\small \{Yuqi.Zhou, Joseph.Severino, Sanjana.Vijayshankar, Juliette.Ugirumurera, Jibo.Sanyal\}@nrel.gov}}%
}

\allowdisplaybreaks

\begin{document}

\maketitle
\thispagestyle{empty}
\pagestyle{empty}

\begin{abstract}
Timely and effective load shedding in power systems is critical for maintaining supply-demand balance and preventing cascading blackouts. To eliminate load shedding bias against specific regions in the system, optimization-based methods are uniquely positioned to help balance between economic and fairness considerations. However, the resulting optimization problem involves complex constraints, which can be time-consuming to solve and thus cannot meet the real-time requirements of load shedding. To tackle this challenge, in this paper we present an efficient machine learning algorithm to enable millisecond-level computation for the optimization-based load shedding problem. Numerical studies on both a 3-bus toy example and a realistic RTS-GMLC system have demonstrated the validity and efficiency of the proposed algorithm for delivering fairness-aware and real-time load shedding decisions.

\end{abstract}


\section{Introduction}
Quick restoration of power balance during emergency events is vital for ensuring stable system operations and preventing system-wide cascading failures. In practice, load shedding serves as a common emergency response employed by system operators to maintain grid stability by selectively shutting off non-critical loads. During severe contingencies such as extreme weather events, timely and effective load shedding actions are critical to prevent grid collapse. However, the execution of load shedding would inevitably cause inconvenience for customers and even raise safety and health concerns. Therefore, it is important to develop a fairness-aware load shedding decision-making model to eliminate bias among different areas, while still meeting the time-sensitive requirements of such operations.

To account for grid-wide impact and pursue the most economical decisions, optimization-based load shedding schemes have become popular. These problems are typically formulated to target either network operational limits (e.g., \cite{coffrin2018relaxations,rhodes2020balancing,zhou2025machine}) or system power balance (e.g., \cite{larik2018improved,xu2011stable}). More recently, fairness concerns have been investigated for inclusion in the load shedding decision-making framework. For example, the vulnerability of end users has been considered in \cite{taylor2023managing} to prevent biased load shedding actions towards specific communities. Similarly, \cite{sundar2023fairly} presents methods to integrate fairness into the minimum load shedding problem, realized using second-order cone constraints. In addition, the fairness of load shedding under a rolling-horizon setup have also been studied in \cite{kody2022sharing,zhou2024equitable}. However, the majority of these works rely on relaxation techniques to enhance the computational efficiency of the optimization problem, but still cannot deliver the real-time computation necessary to meet the extremely time-critical requirements (typically within milliseconds to seconds) of the load shedding decision-making.

In this paper, we aim to develop an efficient machine learning algorithm to enable real-time computation of fairness-aware load shedding decisions. To this end, we first formulate an optimization-based load shedding problem. To ensure fairness in the load shedding decision-making, diverse constraints are enforced to discourage disproportionate load shedding and eliminate bias against various features associated with each node or region. After that, we propose utilizing machine learning algorithms to accurately determine the binding or non-binding status of each inequality constraint based on different inputs. Once the binding constraints are identified, the original problem can be simplified to an equality-constrained optimization. By analyzing the optimality conditions, the optimal solutions to the original problem can be obtained by simply solving a set of linear equations. Instead of incorporating all the optimization constraints, the learning algorithm allows us to identify only the critical constraints in advance to the actual computation. These binding patterns can then be effectively used to establish a linear system, to facilitate the real-time computation of load shedding decisions.

This paper is organized as follows. Section \ref{sec:els} introduces the optimization formulation of the load shedding problem. Section \ref{sec:learning} focuses on the optimality analysis as well as the machine learning algorithm. Section \ref{sec:ns} uses both a small system and an RTS-GMLC system to demonstrate the validity and computational efficiency of the proposed algorithm. Lastly, Section \ref{sec:con} concludes the paper.

\textit{Notation:}  Upper- (lower-) case boldface symbols are used to denote matrices (vectors); $(\cdot)^{\mathsf T}$ stands for transposition; $\mathbf{0}$ and $\mathbf{1}$ denote the all-zero and all-one vectors; $\mathds{1}$ denotes the indicator function; $\mathbf{e}_i$ denotes the standard basis vector with all entries being 0, except for the $i$-th entry being 1; $(\cdot)^{*}$ denotes the optimal value of a decision variable; $\mathbb{E}(\cdot)$ denotes the expected value of a random variable; $\text{ker}(\cdot)$ denotes the kernel of a matrix.


\section{Fairness-Aware Load Shedding Problem}
\label{sec:els}

Consider a power network with a total of $N$ buses collected in the set $\cal N :=$ $\{1,\ldots,N\}$ and $L$ lines in the set $\cal L :=$ $\{l = (i,j)\} \subset \cal N \times \cal N$. For each bus $i$, let $\theta_i$ denote its phase angle and collect all the angles in $\bm{\theta} \in \mathbb{R}^{N}$. Similarly, we use vectors $\bm{g} \in \mathbb{R}^{N}$, $\bm{d} \in \mathbb{R}^{N}$ to denote generation and load at all buses. Under the DC power flow \cite{stott2009dc} model, the line flows $\{f_{ij}\}$, represented by the vector $\bm{f} \in \mathbb{R}^L$, are given as follows:
\begin{align}
\bm{f} = \mathbf{K}\bm{\theta} \label{eq:PF1}
\end{align}
where the matrix $\mathbf{K} \in \mathbb R^{L\times N}$ serves to map the phase angles of the buses to the line flows. The row of $\mathbf K$ associated with line $(i,j)$ can be given by $b_{ij} (\mathbf{e}_i-\mathbf{e}_j)^\mathsf T$, where $b_{ij}$ represents the inverse of line reactance. In addition, the following linear equation holds in accordance with the nodal power balance:
\begin{align}
\bm{p} = \mathbf{A}\bm{f} \label{eq:PF2}
\end{align}
in which $\bm{p}$ is the vector of net injection, and $\mathbf{A} \in \mathbb{Z}^{N \times L}$ is the incidence matrix for the underlying graph $(\cal N, \cal L)$.

During power system operations, load shedding is required occasionally to balance supply and demand, preventing system collapse and potential cascading failures. Optimization-based load shedding schemes can help accurately determine the most cost-effective solutions, but the corresponding decisions may exhibit biases across different locations. Such biases are even more likely to occur during emergency operations or in situations where load shedding is required to relieve local congestion. Therefore, it is crucial to account for various features (e.g., age, income, housing type, access to resources, etc) associated with each area when designing the load shedding algorithm to mitigate bias. Consider a collection of normalized vectors $\bm{v}^{k} \ (k = 1, \cdots, K)$, each representing a unique feature across all nodes. Our goal is to determine the load shedding percentage $s_{i} \in [0,1]$ for each node $i \in \cal N$ to promote fairness in decision-making. Hence, the fairness-aware load shedding problem can be formulated as follows:
\begin{subequations}\label{eq:LS}
\begin{align} 
\min_{\bm{\theta}, \bm{g},\bm{f},\bm{s}} \quad &  \sum_{i \in \cal N} \left(a^{2}_{i} g^{2}_{i} + b_{i} g_{i} + c_{i} \right) + \lambda \sum_{i \in \cal N} s_{i} d_{i}
   \label{eq:LS_a}\\
\textrm{s.t.} \quad  
  &  f_{\ell} = b_{ij} (\theta_i - \theta_j), \; \forall \ell = (i,j) \label{eq:LS_b}\\
  & \sum_{\ell : i\in \mathcal N_\ell} f_{l} = {g}_{i} - (1-s_{i}){d}_{i},  \; \forall i \label{eq:LS_c}\\
  & {\theta}_{i}^{\min} \leq {\theta}_{i} \leq{\theta}_{i}^{\max},  \; \forall i \label{eq:LS_d}\\
  & {g}_{i}^{\min} \leq {g}_{i} \leq {g}_{i}^{\max},  \; \forall i    \label{eq:LS_e}\\
  & {f}_{\ell}^{\min}  \leq {f}_{\ell} \leq {f}_{\ell}^{\max},  \; \forall \ell \label{eq:LS_f}\\
  & 0 \leq s_{i} \leq s^{\max}_{i} \leq 1,  \; \forall i \label{eq:LS_g}\\
  & s_i \leq \frac{\gamma}{N} \cdot \left(\sum_{i=1}^{N} s_{i}\right), \; \forall i \label{eq:LS_h}\\
  & \left|s_i - s_j\right| \leq \delta_{ij}, \; \forall (i,j) \label{eq:LS_i}\\
  & \bm{s}^{\mathsf T} \bm{v}^{k}  \leq \epsilon,  \; \forall k = 1,2,\cdots,K. \label{eq:LS_j}
\end{align} 
\end{subequations}
which constitutes a quadratic program (QP) that can be efficiently solved using optimization solvers such as Gurobi, CPLEX, and MOSEK. The objective includes minimizing the total generation costs, which are represented by a quadratic function of $\bm{g}$, in which $a_{i}, b_{i}, c_{i}$ denote the quadratic, linear, and constant coefficients. In addition, the objective includes the total load shedding $\sum_{i \in \cal N} s_{i} d_{i}$, where $s_{i}$ specifies the load shedding percentage for each node $i$. A coefficient $\lambda$ is used to regulate the weight of the load shedding term in the objective function. In particular, we set the parameter $\lambda$ using a sufficiently large number, to ensure the minimization of total load shedding is prioritized over the consideration of generation costs.

The decision variables include phase angles $\bm{\theta}$, generation $\bm{g}$, line flow $\bm{f}$, and load shedding $\bm{s}$. Equality constraints \eqref{eq:LS_b} and \eqref{eq:LS_c} correspond to DC power flow and nodal power balance, per \eqref{eq:PF1} and \eqref{eq:PF2}. The operational limits for the decision variables are specified in constraints \eqref{eq:LS_d} - \eqref{eq:LS_g}, respectively. Notably, the upper limit of load shedding percentage $s^{\max}_{i}$ can be adjusted to less than $1$ to accommodate non-flexible critical loads. Constraint \eqref{eq:LS_h} is enforced to ensure none of the load shedding is disproportionately high across the system. The parameter $\gamma$ is chosen within the interval of $(1, N]$, where a smaller $\gamma$ value helps to reduce bias and promote fairness of load shedding among all loads. Additionally, constraint \eqref{eq:LS_i} is introduced to limit the discrepancy in load shedding of any pair $(i,j)$, with a smaller $\delta$ generally promoting more fair decisions. To further mitigate biased decisions under normalized features $\bm{0} \leq \bm{v}^{k} \leq \bm{1}$, the constraint \eqref{eq:LS_j} is enforced. The inner product $\bm{s}^{\mathsf T} \bm{v}^{k}$ is constrained within $\epsilon$, to encourage the load shedding decisions to be independent of each unique feature. A small $\epsilon$ value encourages the decision vector to be orthogonal to these features to enhance fairness, while larger $\epsilon$ values relax the constraint and allow for greater flexibility in load shedding decisions under these features. 


\begin{remark}[Notation of index]
With slight abuse of notation, the nodal index $i \in \cal N$ is also used for generation, load, and load shedding in \eqref{eq:LS}. In scenarios where not all nodes are equipped with generation or load, the sets can be altered to $\cal G$ and $\cal D$, with moderate modifications needed for the notations and formulations in \eqref{eq:LS}.
\end{remark}

\begin{remark}[AC load shedding formulation]
In this work, our primary focus is on exploring the learning approach to the load shedding problem, which involves solving it repetitively to collect enough data inputs/outputs. For simplicity, we adopt the DC power flow-based optimization formulation in \eqref{eq:LS}. A nonlinear formulation based on the AC power flow model is also possible (see e.g., \cite{coffrin2018relaxations,larik2018improved}). 
\end{remark}

While the QP problem \eqref{eq:LS} can be solved effectively using state-of-the-art optimization solvers, we are interested in investigating the learning-to-optimize approach in this research. The reasons for this are threefold:
\begin{itemize}
  \item The decision-making for load shedding during emergency operations is highly time-sensitive, and thus developing an efficient learning algorithm can facilitate online solutions for real-time emergency responses.
  \item Instead of approaching the load shedding merely as a generalized optimization problem, the learning method also provides a deeper understanding of the specific impact of the cost function, congestion, operational capabilities, and other constraints on the decisions.
  \item The actual operations of load shedding may involve time-varying features and uncertain loads, and machine learning can effectively address these concerns in an offline fashion to enable robust online decision-making under uncertainty.
\end{itemize}

\section{Learning for Real-time Load Shedding}
\label{sec:learning}

For simplicity in the math expressions, let us first collect all the decision variables of problem \eqref{eq:LS} in a vector $\bm x$ such that $\bm{x}^{\mathsf T} = \left[\bm{\theta}^{\mathsf T} \, \bm{g}^{\mathsf T} \, \bm{f}^{\mathsf T} \, \bm{s}^{\mathsf T}\right]$.
Hence, the load shedding problem can be cast as the following QP in the generalized form:
\begin{subequations}
\label{eq:original_QP}
\begin{align}
\llap{\textbf{(original)} \quad} \min_{{{\bm x}}} \quad & \frac{1}{2} {\bm x}^{\mathsf T}{\bm P}{\bm x} + {\bm q}^{\mathsf T}{\bm x} + r \\
\textrm{s.t.} \quad \: 
  & {\bm A} {\bm x} \leq {\bm b} \label{eq:original_QP_b}\\
  & {\bm G} {\bm x} = {\bm h} \label{eq:original_QP_c}
\end{align}
\end{subequations}
where $\bm P$ is a positive semidefinite matrix, $\bm q$ is a vector of linear coefficients, and $r$ represents the constant term. The inequality \eqref{eq:original_QP_b} match with \eqref{eq:LS_d} - \eqref{eq:LS_j}, and the equality \eqref{eq:original_QP_c} correspond to \eqref{eq:LS_b} and \eqref{eq:LS_c} in problem \eqref{eq:LS}.

Suppose a subset of inputs $\boldsymbol{\pi}$ to the optimization problem consistently lead to the same binding (tight) constraints ${\bm A}_{\boldsymbol\tau} {\bm x} = {\bm b}_{\boldsymbol\tau}$, then the remaining non-binding constraints can be eliminated as they do not ``actively'' restrict the problem or affect the solution (see also \cite{baker2019joint,misra2022learning,bertsimas2022online}). Hence, computing the optimal solution to the original problem \eqref{eq:original_QP} is also equivalent to solving the following equality-constrained QP problem:
\begin{subequations} \label{eq:reduced}
\begin{align} 
\llap{\textbf{(equivalent)} \:} \min_{{{\bm x}}} \quad & \frac{1}{2} {\bm x}^{\mathsf T}{\bm P}{\bm x} + {\bm q}^{\mathsf T}{\bm x} + r\\
\textrm{s.t.}  \quad  
  & {\bm A}_{\boldsymbol\tau(\boldsymbol\pi)} {\bm x} = {\bm b}_{\boldsymbol\tau(\boldsymbol{\pi})} \label{eq:reduced_b}\\
  & {\bm G} {\bm x} = {\bm h} \label{eq:reduced_c}
\end{align}
\end{subequations}
Here, the binding constraints at optimality $\boldsymbol\tau(\boldsymbol\pi)$ are given as a function of input parameters $\boldsymbol{\pi}$. Machine learning methods can be leveraged to effectively learn the binding constraints using inputs $\boldsymbol{\pi}$. Once the binding constraints are determined, the Karush-Kuhn-Tucker (KKT) conditions for this equality-constrained optimization problem can be given as:
\begin{align} \label{eq:optimality}
& {\bm P}{\bm x}^{*} + {\bm q} + {\bm A}^{\mathsf T}_{\boldsymbol\tau(\boldsymbol\pi)} {\bm{v}}^{*} + {\bm G}^{\mathsf T} {\bm w}^{*} = {\bm 0}\\
& {\bm A}_{\boldsymbol\tau(\boldsymbol\pi)} {\bm x}^{*} = {\bm b}_{\boldsymbol\tau(\boldsymbol\pi)}\\
& {\bm G} {\bm x}^{*} = {\bm h}
\end{align}
The vectors ${\bm v}^{*}$ and ${\bm w}^{*}$ denote dual variables associated with constraints \eqref{eq:reduced_b} and \eqref{eq:reduced_c}, respectively. Equivalently, these conditions can be written as a linear system:
\begin{align} \label{eq:kkt}
\begin{bmatrix} 
{\bm P} & {\bm A}^{\mathsf T}_{\boldsymbol\tau(\boldsymbol\pi)} & {\bm G}^{\mathsf T} \\
{\bm A}_{\boldsymbol\tau(\boldsymbol\pi)} & {\bm 0} & {\bm 0} \\
{\bm G} & {\bm 0} & {\bm 0}
\end{bmatrix} 
\begin{bmatrix} 
{\bm x}^{*} \\
{\bm v}^{*} \\
{\bm w}^{*}
\end{bmatrix} = 
\begin{bmatrix} 
-{\bm q} \\
{\bm b}_{\boldsymbol\tau(\boldsymbol\pi)} \\
{\bm h}
\end{bmatrix}
\end{align}
As the KKT matrix on the left-hand side of equation \eqref{eq:kkt} is a square matrix, the solvability of \eqref{eq:kkt} is primarily determined by whether the KKT matrix is singular or nonsingular.


\begin{corollary}
Given that the original load shedding problem \eqref{eq:original_QP} has a unique optimal solution, the linear system \eqref{eq:kkt} is solvable and the KKT matrix is nonsingular.
\end{corollary}

On the other hand, the nonsingularity of the KKT matrix guarantees its invertibility and the existence of unique optimal primal-dual solutions $\{{\bm x}^{*},{\bm v}^{*},{\bm w}^{*}\}$. According to the equivalent conditions \cite[Ch. 10.1]{boyd2004convex} of nonsingularity of KKT matrix, it is nonsingular if the block matrices have no nontrivial common nullspace:
\begin{align}  \label{eq:singularity}
    {\text{ker}} (\bm{P}) \cap {\text{ker}} \left(  \begin{bmatrix} 
    {\bm A}_{\boldsymbol\tau(\boldsymbol\pi)}\\
    \bm{G}\\
    \end{bmatrix} \right) = \{ {\mathbf{0}} \}
\end{align}
Recall that $\bm P$ is a sparse matrix with $2a_i$ on the diagonals. Meanwhile, as long as the coefficient matrix of equality constraints \eqref{eq:reduced_b} and \eqref{eq:reduced_c} is full rank, the nullspace (or kernel) of it is the zero vector. Accordingly, the two block matrices in \eqref{eq:singularity} share only the trivial nullspace $\{\mathbf{0}\}$, and the KKT matrix is nonsingular.
Although one can compute the solution to the linear system by directly applying the inverse of the KKT matrix to both sides of equation \eqref{eq:kkt}, this method may not be the most efficient. Because the KKT matrix is both sparse and symmetric, it fits well for direct solvers \cite{davis2006direct} that use matrix factorization techniques (e.g., LU decomposition). In addition, in cases where the system size is large, indirect methods \cite{paige1975solution} (e.g., MINRES, SYMMLQ) can also be used to facilitate online computation for making real-time decisions.

\begin{figure}[t!]
\centering
\includegraphics[trim=0cm 0cm 0cm 0cm,clip=true,totalheight=0.18\textheight]{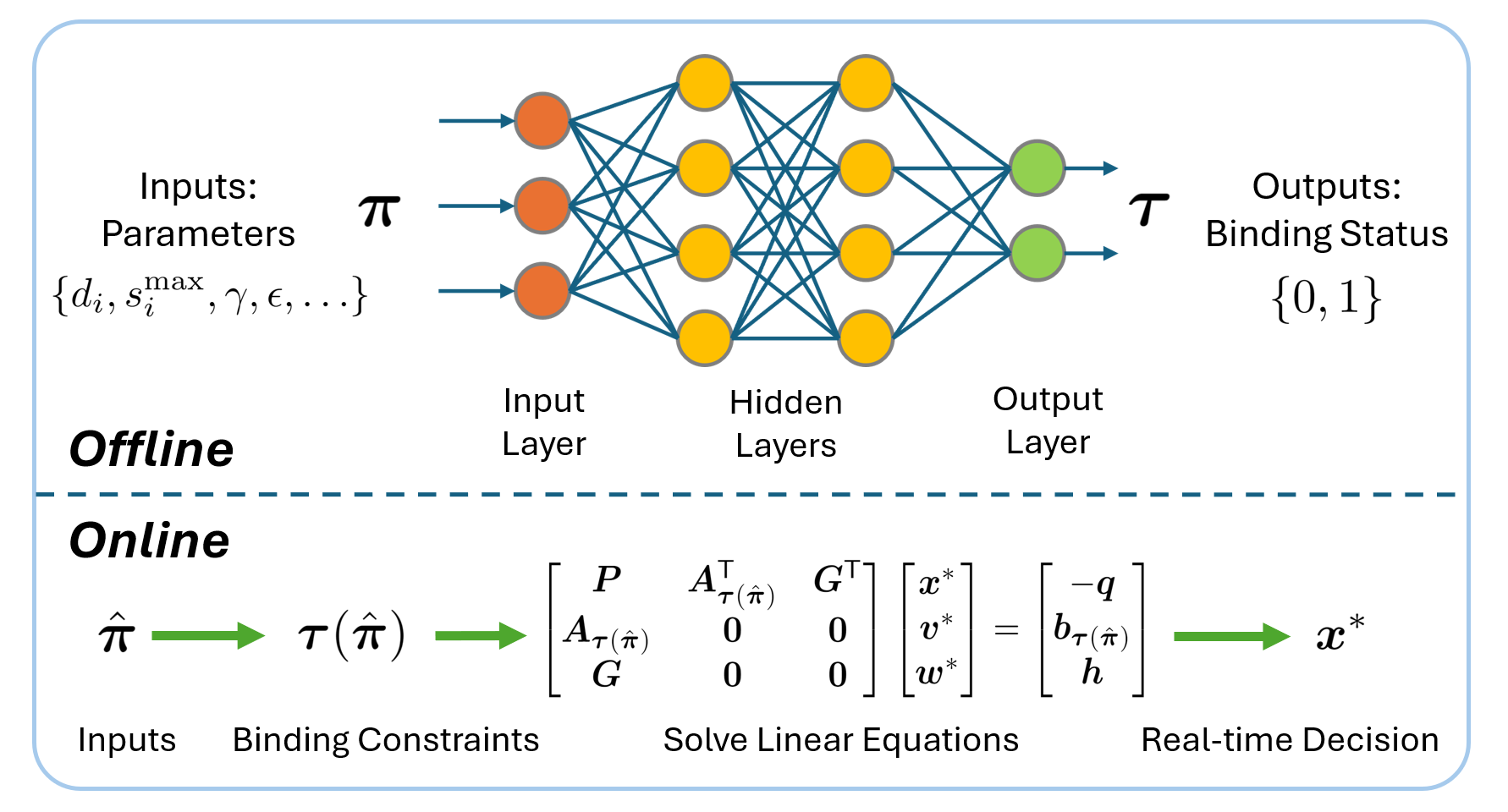}
\caption{The overall framework for learning the fairness-aware load shedding problem to enable real-time decision-making.}
\label{fig:framework}
\vspace{-2mm}
\end{figure}

Clearly, solving the linear system \eqref{eq:kkt} is much more efficient than directly tackling the original QP problem \eqref{eq:original_QP}. However, it necessitates accurately learning the binding constraints offline to facilitate online computation. To identify binding or non-binding constraints, it typically involves the use of binary outputs. 

\begin{remark}[Choice of the machine learning algorithm]
Although many classification algorithms (e.g., Random Forest, Naive Bayes, Support Vector Machine) are available for learning the binding constraints, in this work we adopt a generalized deep neural network (DNN) model. This is because it is efficient in capturing nonlinear relationships, offering higher flexibility \cite{pan2022deepopf} in model configuration and better scalability \cite{zhou2022scalable} with large datasets.
\end{remark}

While the inputs $\boldsymbol{\pi}$ of the learning model could represent any parameters associated with inequality constraints in problem \eqref{eq:LS}, our focus in this work is primarily on using load as inputs. The outputs $\boldsymbol\tau$ constitute a vector of binary numbers to indicate the binding status of each constraint, where $1$ represents binding and $0$ is for non-binding. In addition to verification through a comparison of optimal solutions and their respective inequality bounds, the binding status can be also obtained using the optimal dual solutions.

\begin{proposition}[Binding status via dual solutions] \label{eq:prop1}
For any inequality constraint in \eqref{eq:original_QP_b}, the complementary slackness condition \cite[Ch. 5.5]{boyd2004convex} states that:
\begin{align}
    \mu_i (\bm{A}_{i} \bm{x} - {b}_{i}) = 0, \: \: \forall i
\end{align}
where $\bm{A}_{i}$ denotes the $i$-th row of the coefficient matrix $\bm{A}$.
Hence, if a dual variable $\mu_i = 0$, the related constraint is non-binding or does not affect the optimal objective. On the other hand, if $\mu_i > 0$, the constraint has to be binding ($\bm{A}_{i} \bm{x} = {b}_{i}$). To that end, the binding status of each inequality constraint can be determined by:
\begin{align}
    \tau_i = \mathds{1}_{\{\mu_i > 0\}} = 
\begin{cases} 
1, & \textup{if } \mu_i > 0 \\
0, & \textup{if } \mu_i = 0 
\end{cases}
\end{align}
which is a simple, single-step transformation once the original optimization problem is solved.
\end{proposition} 
The relationship between the input and output across $M$ layers in the neural network can be represented by a set of equations:
\begin{align} \label{eq:nn}
    {\bm y}^{m} = \sigma^{m}\left(\bm{W}^{m} {\bm y}^{m-1} + {\bm b}^{m}\right), \: \: \forall m = 1, \ldots, M
\end{align}
where the first layer is ${\bm y}^{0} = \boldsymbol{\pi}$ and the last layer is ${\bm y}^{M} = \boldsymbol{\tau}$. In the $m$-th layer, the weight matrix and bias vector are denoted by $\bm{W}^{m}$ and $\bm{b}^{m}$, and $\sigma^{m}$ represents the activation function (e.g., ReLU, Sigmoid). The binary cross-entropy is a widely adopted loss function for binary classification tasks:
\begin{align}
    \mathcal{L} = -\frac{1}{J} \sum_{i=1}^J \left[ y_i \log(\hat{y}_i) + (1 - y_i) \log(1 - \hat{y}_i) \right]
\end{align}
Here, $J$ is the number of inequality constraints to learn, $y_i$ is the true label for the $i$-th constraint, and $\hat{y}_i$ is the predicted probability that the constraint is binding. In scenarios where one class dominates the classification of binding and non-binding constraints, the loss function can be adjusted to focal loss to address class imbalance issues.

It is important to realize that our analysis to this point assumes a known load vector. However, due to the uncertainty of load, actual decision-making may require consideration under a broader range of load configurations \cite{chen2019probabilistic}. For reliable load shedding operations, it is generally safer to assume larger loads when making decisions. By and large, assuming smaller loads is more risky, as the resulting load shedding is often limited and may be insufficient to recover the power balance. Accordingly, utilizing risk measures like Value at Risk (VaR) and Conditional Value at Risk (CVaR) allows for flexible adjustment of load inputs to match different levels of risk. Assume that each load $d_{i}$ is a random variable with cumulative distribution function $F_{d_i}(z) = P\{d_i \leq z\}$. The VaR of $d_i$ can be defined as:
\begin{align}
    \text{VaR}_{\alpha}(d_i) = \min\{z|F_{d_i}(z) \geq \alpha\}, \ \forall i \in \cal{N}
\end{align}
where $\alpha \in (0,1)$ is the confidence level (e.g. $\alpha = 0.95$). Another risk measure is CVaR, which is designed to calculate the expected loss in the tail of distribution beyond VaR:
\begin{align}
    \text{CVaR}_{\alpha}(d_i) = \mathbb{E}[d_i \mid d_i \geq \text{VaR}_{\alpha}(d_i)], \ \forall i \in \cal{N}
\end{align}
In many applications, CVaR \cite{rockafellar2000optimization} is preferred over VaR as it is a more coherent risk measure and does not encourage risk-taking that could be obscured by the measure itself. Additionally, it relies less on the underlying distribution and provides a more reliable measure for skewed distributions.

\begin{proposition}[Risk-averse load shedding] \label{eq:prop2} 
To transition from the original risk-neutral (e.g., using predicted load values) to the risk-averse load shedding, one can use the CVaR of the load distribution at each node:
\begin{align}
{\hat{d}}_{i} = \text{CVaR}_{\alpha_i}({d}_i), \ \forall i \in \cal{N}
\end{align}
where ${\hat{d}}_{i}$ denotes the risk-averse load value, and $\alpha_{i}$ allows for flexible adjustment of risk levels associated with each load. Thus, the collection of loads:
\begin{align}
    {\hat{\bm d}} = \{{\hat{d}}_{1}, {\hat{d}}_{2}, \cdots, {\hat{d}}_{N}\}
\end{align}
can serve as the input to obtain $\boldsymbol\tau$ in \eqref{eq:kkt}. Suppose each load is constrained within $d^{\min}_{i} \leq d_i \leq d^{\max}_i, \forall i \in \mathcal{N}$, then it is also possible to pursue robust solutions by using:
\begin{align}
    {\hat{\bm d}_{\text{r}}} = \{{d}^{\max}_{1}, {d}^{\max}_{2}, \cdots, {d}^{\max}_{N}\}
\end{align}
to ensure robustness against the worst-case scenario.
\end{proposition}








\section{Numerical Simulation}
\label{sec:ns}

In this section, we present the numerical simulation results using a 3-bus toy example and a 73-bus RTS-GMLC system with realistic datasets. The fairness-aware load shedding problem is implemented with MATPOWER, and solved using the Gurobi solver. The machine learning algorithm is implemented in MATLAB with the deep learning toolbox. The simulations are performed on a regular laptop with Intel\textsuperscript{\textregistered} CPU @ 2.60 GHz and 16 GB of RAM.

\subsection{Illustrative Example with a 3-Bus System}
\begin{figure}[t!]
\centering
\includegraphics[trim=0cm 0.8cm 1.8cm 0cm,clip=true,totalheight=0.14\textheight]{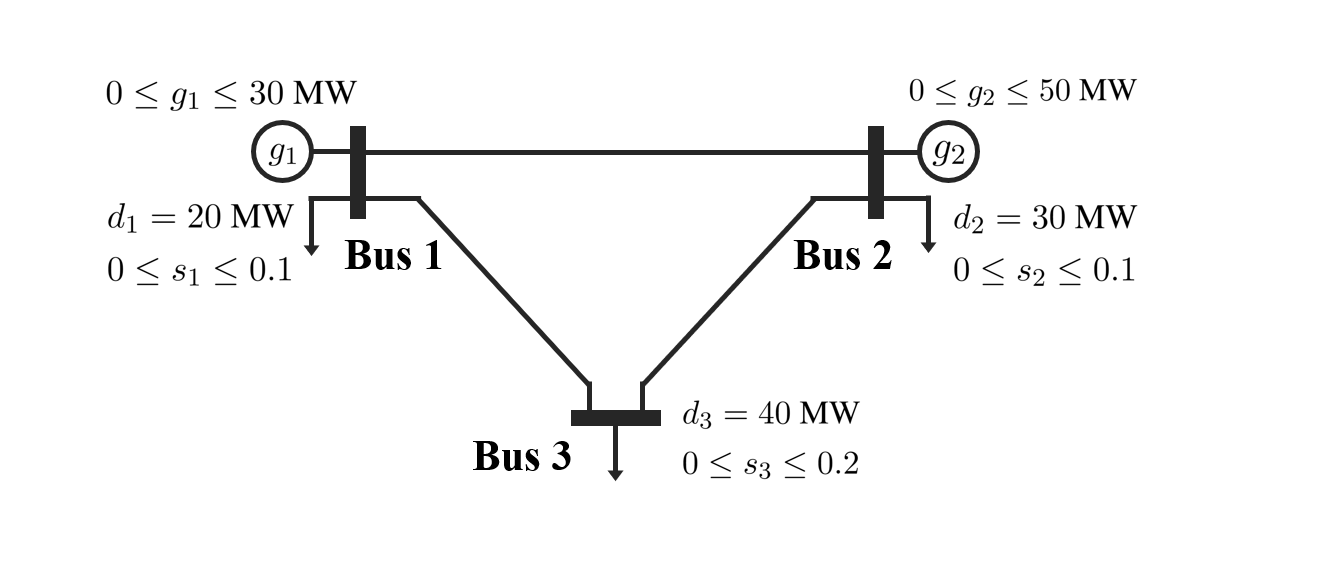}
\caption{A 3-bus illustrative example with two generators.}
\label{fig:3_bus_figure}
\end{figure}

We begin with a small 3-bus system to verify the optimality condition and demonstrate how to efficiently handle the optimization problem by solving simple linear equations. The 3-bus system has two generators and three loads. We use the quadratic cost function $c(g_1) = g^{2}_{1} + 3g_{1}$ and $c(g_2) = 2g^{2}_{2} + g_{2}$, for each generator. The generation limits, load values, and load shedding limits are provided in Figure.~\ref{fig:3_bus_figure}. The test case is designed so that the total load exceeds the generation capacity, which necessitates the load shedding operations. Furthermore, we have simplified the problem by removing the transmission constraint between any pair of buses. To this end, we can formulate the load shedding problem as follows:
\begin{subequations}
\label{eq:3_bus}
\begin{align}
\min \quad & (g^{2}_{1} + 3g_{1} + 2g^{2}_{2} + g_{2}) + \lambda\left(20s_{1}+30s_{2}+40s_{3}\right) \nonumber\\
\textrm{s.t.} \quad 
  & 0 \leq g_{1} \leq 30, 0 \leq g_{2} \leq 50   \nonumber\\
  & 0 \leq s_{1} \leq 0.1, 0 \leq s_{2} \leq 0.1, 0 \leq s_{3} \leq 0.2  \nonumber\\
  & s_{i} \leq 0.5(s_1+s_2+s_3), \: \forall i = 1,2,3  \nonumber\\
  & g_1 + g_2 = 20(1-s_1) + 30(1-s_2) + 40(1-s_3) \nonumber
\end{align}
\end{subequations}
where the problem enforces limits for generation and load shedding, as well as a constraint for power balance. In addition, a simple fairness constraint is enforced in this example to ensure that the load shedding at each bus does not exceed half of the total load shedding values. By solving the above optimization problem and eliminating all non-binding constraints, we arrive at the following equivalent problem:
\begin{subequations}
\label{eq:3_bus_equivalent}
\begin{align}
\min \quad & (g^{2}_{1} + 3g_{1} + 2g^{2}_{2} + g_{2}) + \lambda\left(20s_{1}+30s_{2}+40s_{3}\right) \nonumber\\
\textrm{s.t.} \quad 
  & g_{1} = 30 \quad (v_1), \quad g_{2} = 50 \quad (v_2)   \nonumber\\
  & s_{1} = 0.1 \quad (v_3), \quad s_{2} = 0.1 \quad (v_4)   \nonumber\\
  & g_1 + g_2 + 20s_1 + 30s_2 + 40s_3 = 90 \quad (w_1) \nonumber
\end{align}
\end{subequations}
Let $v_1, v_2, v_3, v_4, w_1$ denote the dual variables of each equality constraint above. Accordingly, the KKT conditions can be equivalently expressed as the following linear system:
\begin{align}
\left[
\begin{array}{c@{\hspace{0.5em}}c@{\hspace{0.5em}}c@{\hspace{0.3em}}c@{\hspace{0.3em}}c|@{\hspace{0.3em}}c@{\hspace{0.4em}}c@{\hspace{0.4em}}c@{\hspace{0.4em}}c|@{\hspace{0.4em}}c}
2 & 0 & 0 & 0 & 0 & 1 & 0 & 0 & 0 & 1 \\
0 & 4 & 0 & 0 & 0 & 0 & 1 & 0 & 0 & 1 \\
0 & 0 & 0 & 0 & 0 & 0 & 0 & 1 & 0 & 20 \\
0 & 0 & 0 & 0 & 0 & 0 & 0 & 0 & 1 & 30 \\
0 & 0 & 0 & 0 & 0 & 0 & 0 & 0 & 0 & 40 \\ \hline
1 & 0 & 0 & 0 & 0 & 0 & 0 & 0 & 0 & 0 \\
0 & 1 & 0 & 0 & 0 & 0 & 0 & 0 & 0 & 0 \\
0 & 0 & 1 & 0 & 0 & 0 & 0 & 0 & 0 & 0 \\
0 & 0 & 0 & 1 & 0 & 0 & 0 & 0 & 0 & 0 \\ \hline
1 & 1 & 20 & 30 & 40 & 0 & 0 & 0 & 0 & 0 \\
\end{array}
\right] \begin{bmatrix} 
g_1\\
g_2\\
s_1\\
s_2\\
s_3\\
v_1\\
v_2\\
v_3\\
v_4\\
w_1
\end{bmatrix} = 
\begin{bmatrix} 
-3\\
-1\\
-20\lambda\\
-30\lambda\\
-40\lambda\\
30\\
50\\
0.1\\
0.1\\
90
\end{bmatrix} \nonumber
\end{align}
In this example, we can set the parameter to be $\lambda = 1000$, to prioritize traditional generation over load shedding. By solving the optimization problem and the linear system, we arrive at the exact same \textbf{primal solutions} $g^{*}_1 = 30$, $g^{*}_2 = 50$, $s^{*}_1 = 0.1$, $s^{*}_2 = 0.1$, $s^{*}_3 = 0.1$ and the \textbf{dual solutions} $v^{*}_1 = 937$, $v^{*}_2 = 799$, $v^{*}_3 = 0$, $v^{*}_4 = 0$, $w^{*}_1 = -1000$.
Suppose the machine learning algorithm identifies that every load configuration  $\{d_1,d_2,d_3\}$ in a set $S$ results in identical binding constraints for the problem above, then the optimal solutions can be directly obtained by solving the following:
\begin{align}
\begin{bmatrix} 
g_1^{*}\\
g_2^{*}\\
s_1^{*}\\
s_2^{*}\\
s_3^{*}\\
v_1^{*}\\
v_2^{*}\\
v_3^{*}\\
v_4^{*}\\
w_1^{*}
\end{bmatrix} = \left[\begin{array}{c@{\hspace{0.5em}}c@{\hspace{0.5em}}c@{\hspace{0.3em}}c@{\hspace{0.3em}}c|@{\hspace{0.3em}}c@{\hspace{0.4em}}c@{\hspace{0.4em}}c@{\hspace{0.4em}}c|@{\hspace{0.4em}}c}
2 & 0 & 0 & 0 & 0 & 1 & 0 & 0 & 0 & 1 \\
0 & 4 & 0 & 0 & 0 & 0 & 1 & 0 & 0 & 1 \\
0 & 0 & 0 & 0 & 0 & 0 & 0 & 1 & 0 & d_1 \\
0 & 0 & 0 & 0 & 0 & 0 & 0 & 0 & 1 & d_2 \\
0 & 0 & 0 & 0 & 0 & 0 & 0 & 0 & 0 & d_3 \\ \hline
1 & 0 & 0 & 0 & 0 & 0 & 0 & 0 & 0 & 0 \\
0 & 1 & 0 & 0 & 0 & 0 & 0 & 0 & 0 & 0 \\
0 & 0 & 1 & 0 & 0 & 0 & 0 & 0 & 0 & 0 \\
0 & 0 & 0 & 1 & 0 & 0 & 0 & 0 & 0 & 0 \\ \hline
1 & 1 & d_1 & d_2 & d_3 & 0 & 0 & 0 & 0 & 0 \\
\end{array}
\right]^{-1}
\begin{bmatrix} 
-3\\
-1\\
-d_1\lambda\\
-d_2\lambda\\
-d_3\lambda\\
30\\
50\\
0.1\\
0.1\\
\Sigma d
\end{bmatrix} \nonumber
\end{align}

In this toy example, we mainly demonstrate how to arrive at the equivalent linear system using KKT conditions once binding constraints are determined. In addition, numerical results confirm the exactness of solutions from the linear system, as compared to the original optimization problem.

\subsection{RTS-GMLC System}

\begin{figure}[t!]
\centering
\includegraphics[trim=0cm 0cm 0cm 0.4cm,clip=true,totalheight=0.12\textheight]{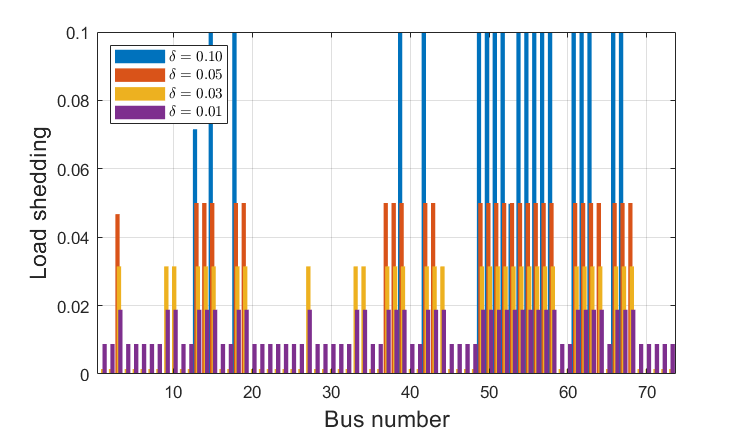}
\caption{Load shedding solutions under varying $\delta$ values.}
\label{fig:GMLC_figure}
\vspace{-0.2cm}
\end{figure}

We next perform numerical simulations on a more realistic RTS-GMLC system \cite{barrows2019ieee}. This synthetic system consists of 73 buses and 120 transmission lines, and covers an area of roughly $250 \times 250$ miles. It models an area in the southwestern United States, roughly extending from Los Angeles to Las Vegas in California. Since the system includes detailed geographic information, we utilize these locations to extract features associated with each node using the communities LEAP \cite{liu2023resstock} data. Each bus in the system is assigned a load, and we extract and normalize 5 nodal features for numerical validation, which are \textbf{\textit{disability, language, minority status, transportation, and housing type}}, respectively. For the load shedding problem \eqref{eq:LS}, we use a quadratic generation cost function and adopt a consistent parameter $\lambda = 10^{4}$ throughout the simulations. An upper limit $s_{i}^{\max} = 0.40$ is set for each bus. Note that for the orthogonality-based fairness constraint, the load shedding decision may not exhibit a direct correlation with the associated threshold, especially in high dimensional space under multiple features. Hence, during the simulation, we set a uniform orthogonality threshold $\epsilon$ for each feature to prevent overly biased decisions. We first investigate the load shedding solutions by varying $\delta$ in \eqref{eq:LS_i}. The load shedding decisions for all buses under different $\delta$ values are given in Fig.~\ref{fig:GMLC_figure}. As this parameter limits the load shedding variations among buses, a smaller $\delta$ promotes more balanced and impartial decisions. Note that pursuing a completely unbiased decision (e.g., $\delta = 0$) may not always be the best choice, as it often leads to uneconomical decisions and may face infeasibility due to the existence of other fairness constraints.

To validate the learning performance and online computation efficiency, we vary the load configurations and solve the optimization repeatedly to collect learning samples. Specifically, we incrementally increase the load (by 5 MW) at buses 5 and 27 to generate $50 \times 50 = 2500$ inputs and solve each for learning outputs. Analysis of the learning samples reveals 7 unique binding status outputs. Among these outputs, the binding status for 6 specific constraints alternates between binding and non-binding. The main causes are the changes in congestion patterns or touches of fairness constraint bounds as load changes. Pre-processing like this can help eliminate outputs with constant binding status, and significantly reduce the output dimension. 
To learn the binding status of the remaining critical constraints, a neural network (NN) is constructed with three hidden layers. The NN is trained using the gradient descent algorithm, with a learning rate of 0.01 and a maximum of 500 epochs.
The tested system achieves an average training time of under 10 seconds, with an overall prediction accuracy of $99.46\%$ for constraints that alternate between binding and non-binding. The small mismatch error is mainly due to how the samples are generated. In particular, as we incrementally adjust the load, the binding pattern can shift from one to another. When inputs are limited, the samples that cover certain binding patterns might be insufficient and thus lead to slightly inaccurate predictions. The issue can be resolved by using smaller increments and more learning samples.
After binding status $\boldsymbol{\tau}$ is determined, linear system \eqref{eq:kkt} is solved to obtain optimal solutions. To show the computational efficiency, we solve both the optimization problem and the linear system multiple times, and record the minimum, maximum, and median computation time in Table~\ref{tab:LR}. Notably, solving linear equations with known binding status is about 20000 times faster than tackling the original problem. Overall, the results demonstrate the effectiveness and accuracy of our learning-to-optimize approach toward real-time load shedding decision-making.

\begin{figure}[t!]
\centering
\includegraphics[trim=0cm 0cm 0cm 0.4cm,clip=true,totalheight=0.12\textheight]{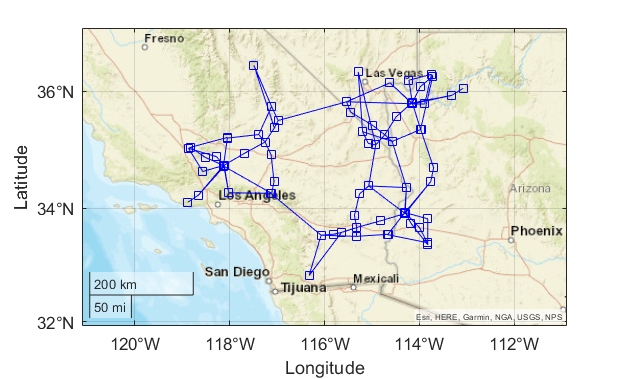}
\caption{Network topology of the RTS-GMLC system.}
\label{fig:GMLC_network}
\vspace{-0.3cm}
\end{figure}

\begin{table}[t!]
\caption{Computation time (in seconds) for solving optimization and linear system, with same optimal solutions.}
\label{tab:LR}
\centering
\renewcommand{\arraystretch}{1.1} 
\begin{tabular}{|p{1.9cm}||p{1.6cm}|p{1.6cm}|p{1.6cm}|}
\hline
& Minimum (s) & Median (s) & Maximum (s) \\
\hline
Optimization & 5.74  & 6.13  & 6.87  \\
\hline
Linear System & 0.00028  & 0.00030  & 0.00035  \\
\hline
Speedup & $2.050 \times 10^{4}$  & $2.043 \times 10^{4}$  & $1.963 \times 10^{4}$  \\
\hline
\end{tabular}
\vspace{-0.5cm}
\end{table}



\section{Conclusions and Future Work}
\label{sec:con}

In this paper, we present an efficient machine learning algorithm for the real-time fairness-aware load shedding problem. The load shedding is formulated as an optimization problem, which aims to eliminate bias among different loads. By learning the binding constraints, the optimization problem can be simplified to solving a set of linear equations. As a result, the load shedding problem can be solved at the millisecond level, making it suitable for real-time operations. Future work includes investigation of efficient learning algorithms to support fairness-aware decision-making in networked distribution systems.




\bibliography{bibliography.bib}





\bibliographystyle{IEEEtran}

\end{document}